# Dispersion-like phenomena in Jovian decametric S-bursts: Tabooed Facts

O.V. Arkhypov and H.O. Rucker

Space Research Institute, Austrian Academy of Sciences, Schmiedlstrasse 6, A-8042 Graz, Austria

**Abstract**

The dominant viewpoint on Jovian decametric S-burst emission neglects the time delay of the radiation, although its base theory of electron cyclotron maser instability allows a significant decreasing of X-mode group velocity near the cutoff frequency at the bottom of source region. We searched for effects of the frequency-related delay of radiation in broadband Jovian radio storms consisting of periodic S-bursts (S-burst trains) at 16 to 30 MHz. It was found that up to 1% of bursts in a train are of distorted meandering shape in dynamic spectrum, where the emission from one radio source was observed at several frequencies simultaneously. It is difficult to explain such spectra in terms of radio waves beaming or causality without significant frequency-related delay of radio emission. We found experimentally that the frequency drift rate of middle lines of such events coincides with the drift rate of disturbances in common S-bursts. This indicates a general distortion of the dynamic spectrum of S-bursts. As a result, the correlation method for the measurement of the spectral distortion is proposed. Using this method, we found the approximation coefficients for the distortion in 32 spectra of 8 Io-B storms. The corrected spectra formally show that S-burst trains do not move mainly outward from Jupiter, as it is usually assumed, but fly in the opposite direction. Our simulation confirms that the dispersion is capable in principle to reproduce the found spectral distortion. Hence, the dispersion-like phenomena in Jovian S-bursts deserve discussion because they have no satisfactory explanations in terms of traditional approach.

## 1. Introduction

Since the pioneering works in the 1960s it is widely accepted that short (S-) bursts of Jovian decametric emission (DAM) drifts in frequency in strict accordance with the movement of their radio sources (e.g., Ellis, 1965; Ryabov, 1994; Hess et al., 2007a). Accordingly, the frequency-dependent delay of the radiation is ignored as a negligible distortion factor in DAM-source dynamics. However, this postulate could not be valid near DAM sources, where the electron cyclotron maser instability generates X-mode near its cut-off frequency $f_x$. The X-mode origin of DAM and its maser generation are a base of the mainstream theory of S-bursts (Zarka, 1998). Moreover, it is well known that X-mode has a small group speed $v_g \ll c$ at $(f - f_x)/f_x \ll 1$, where $c$ is the light velocity, and $f$ is the emission frequency (Hewitt and Melrose, 1983). Willes (2002) argued for the importance of dispersion effect in S-burst phenomenology. Our numerical simulations have shown that the time delay of the radiation is dependent on the frequency, forming the distortion in DAM dynamic spectrum of the same order as the observed frequency drift rate of S-bursts (Arkhypov and Rucker, 2012).

Hence, it is not excluded that DAM dispersion significantly modifies the seeming dynamics and energy of S-burst emitting electrons. In this paper, we try to detect signs of dispersal distortion of common S-bursts in dynamic spectrum using only empirical approach. The emphasis on the empirism is due to the lack of satellite measurements in the region of DAM sources.

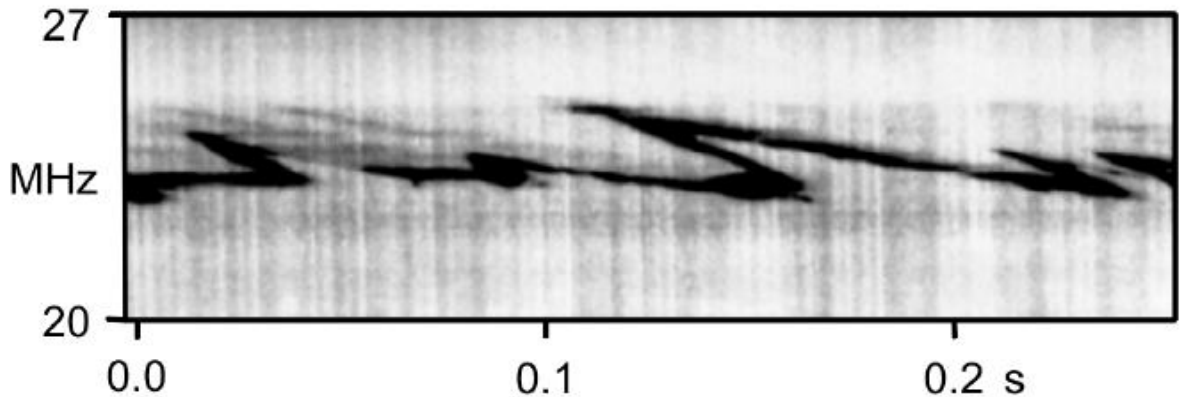

**Figure 1:** *The hooks in the narrow-band emission, which oscillates around average frequency in DAM dynamic spectra, are an argument for a greater time delay of the radio emission at low frequencies. This example is taken from Riihimaa (1992).*

Previously the attention was drawn to the distorted shape of oscillations of narrow-band radiation in DAM dynamic spectra as a manifestation of dispersion (Arkhypov and Rucker, 2011a, 2012). Figure 1 shows the example of such DAM spectrum, where the radiation from one emission band is observed simultaneously at 2 to 3 different frequencies. The only motion of a single radio source in space cannot create this ambiguity in frequency of DAM, because at some time moment the source is located only in one region of space and generates the radiation near one local electron gyrofrequency.

Hitherto such phenomena with ambiguity of the emission frequency were considered heuristically only as a result of correlated motion and generation in groups of 3 radio sources (Oya et al., 2002 and therein; Willes, 2002). Hess et al. (2009), using the non-dispersion simulation, reproduces the frequency-ambiguity effect in common S-bursts as a result of division of the planetward electron beam on passing, reflected and trapped components after interaction with the field of electric potential structure (Hess et al., 2009). However, the common narrow-band oscillations, like in Fig. 1, are an unsolved problem for the non-dispersion approach. The needed number of sources or electric potential structures increases dramatically for multi-periodic events. Such a multi-source constructing seems unnecessarily complicated. We used only one radio-source to reproduce zigzags in S-bursts or narrowband oscillation in dynamic spectrum of DAM, but assuming the existence of a significant dispersion (Arkhypov and Rucker, 2011b).

Another dispersionless alternative is the interpretation of curved S-bursts in terms of narrow beaming of radio waves from extended radio source (Louarn, 1997). However, this approach requires unrealistic narrow radio beams with angle width of $\Omega\tau<2\times10^4$ deg., where $\Omega = 2.3\times10^3$ deg./s is the angle velocity of S-burst source rotation (i.e., the sidereal orbital velocity of related Io satellite); and $\tau<0.1$ s is the time scale of S-bursts. Note that typical experimental estimate of S-burst radio beam is width $\approx1$ deg. (e.g., Ellis, 1982; Zarka, 1998).

Shaposhnikov et al. (2011) proposed the parametric mechanism for the formation of narrow-band oscillations. The hooks-like forms of such oscillations (i.e., the frequency-ambiguity effect) was reproduced using the strong frequency dispersion of X-mode near its cutoff.

Therefore, the significant dispersion of DAM is not excluded. The dispersion-like phenomena could be used for a testing of theoretical models. For example, it is important to study the frequency dependence of the dispersion-like effects. This is interesting, because the non-dispersion modeling of Hess et al. (2009) are based on local effects around electric potential structures (Hess et al., 2009). *Hence, this model predicts that the dispersion-like effect is local and irregularly variable in frequency. In contrary, the dispersion in magnetized plasma is a regular function of frequency.*

That is why our search is focused on broadband (≈10 MHz), quasi-periodical S-bursts (Fig. 2). Traditionally such events are considered as non-dispersed (e.g., Hess et al., 2007a,b). However, it follows from our mathematical modeling that significant dispersion is a required component for reproducing of correlation pattern of S-bursts (Arkhypov and Rucker, 2012). To test these alternatives, we studied the archive records of DAM storms with classical S-bursts.

Our search for symptoms of S-burst distortion and its application are described in Section 2. In Section 3 we correct the dynamic spectra of S-burst storms and analyze them with correlation method. The obtained results are discussed in Section 4. Our conclusions are summarized in Section 5.

## 2. Dispersion of S-burst emission

Besides the classic S-bursts in the form of quasi-linear emission bands in dynamic DAM spectrum (Fig. 2a), there are perturbed bursts of various forms (Fig. 2b). For example, Fig. 3 shows the S-bursts with clear frequency-ambiguity effect: sometimes the radiation of one emission band is seen at 4 frequencies simultaneously. The difficulty in interpreting such curved bursts while ignoring the dispersion of radiation is well demonstrated by Ryabov (2001): "It is difficult to explain these spectra in terms of causality, since the emission occurs at some time independently at the two greatly spaced frequencies, and then these emission bands merge smoothly into a single point on the *f— t* plane". Obviously, the causality problem disappears with the frequency-dependent time-delay of the radiation.

The alternative interpretation in terms of a hypothetical interaction between sources (Willes, 2002) would need of 4 interacting radio sources for the events in Fig. 3. Such complication of the model makes this approach problematic. The assumption of a significant distortion (i.e., dispersion) of DAM dynamic spectrum can easily solve the problem of ambiguity of radiation frequency and to reduce the number of necessary sources to one. For example, the left plots in Fig. 4 show some typical patterns of S-bursts with ambiguity in frequency. However, this ambiguity vanishes when the plots are re-distorted to correct the relative time delay of the emission at low frequencies (right plots in Fig. 4). That is why we adopt the spectral distortion as a working hypothesis for analysis.

In our analysis of spectral distortion, we use *f—t* patterns, like labeled f and h in Fig. 4, which show emission at two frequencies simultaneously. It is important that such elements are found among linear S-bursts (Fig. 5). Their part is up to 1% of usual S-bursts in studied spectra. To eliminate the ambiguity in frequency of such event, we could re-distort the spectrum like in Fig. 4. The problem of frequency ambiguity (i.e., causality problem) disappears in the maximum number of f- and h-events, when their middle lines (medians) are mainly aligned along the frequency axis. Apparently, in this case we obtain the approximation of undistorted spectrum. Hence, we can use the observed inclination of middle lines to estimate the distortional drift rate ($D_d$). The method of such measurements is shown in Fig. 5c:

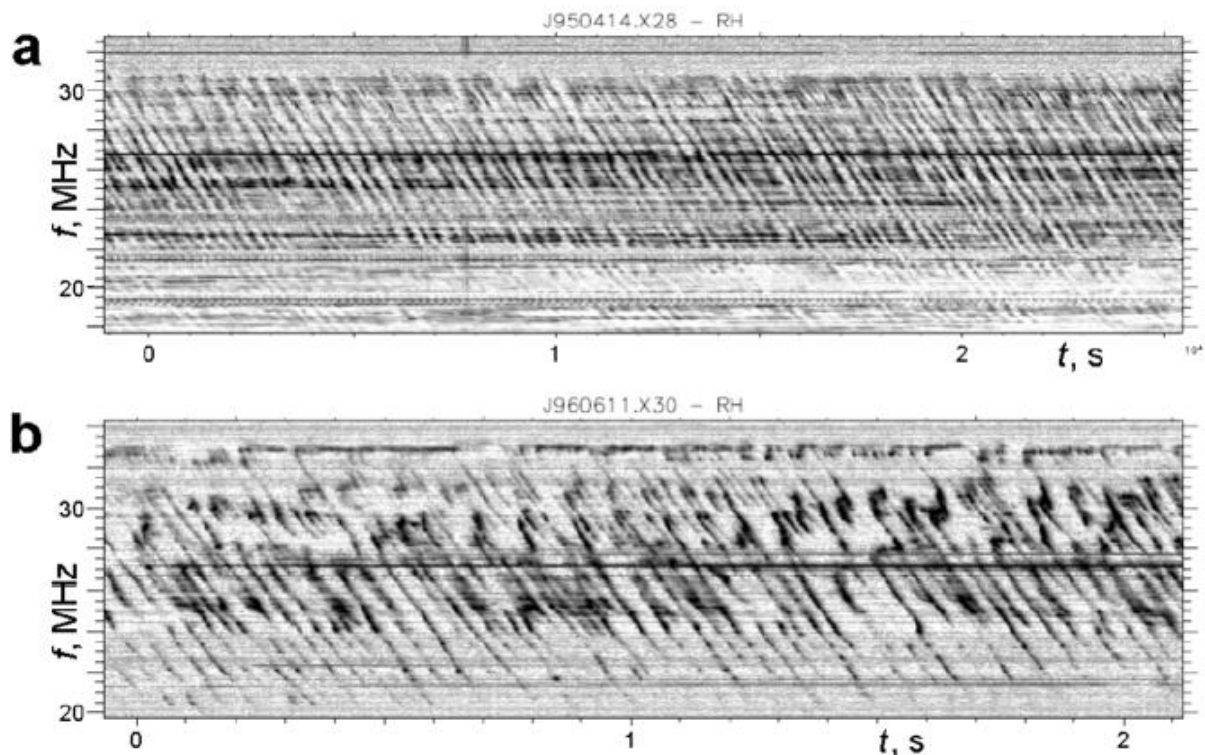

**Figure 2:** *The sample fragments of dynamic spectra where the broadband S-bursts form quasi-periodical trains: (a) the quasi-linear S-bursts (NDA; 1995 April 14, 05:43 UT); (b) the perturbed S-bursts (NDA; 1996 June 11, 23:40 UT).*

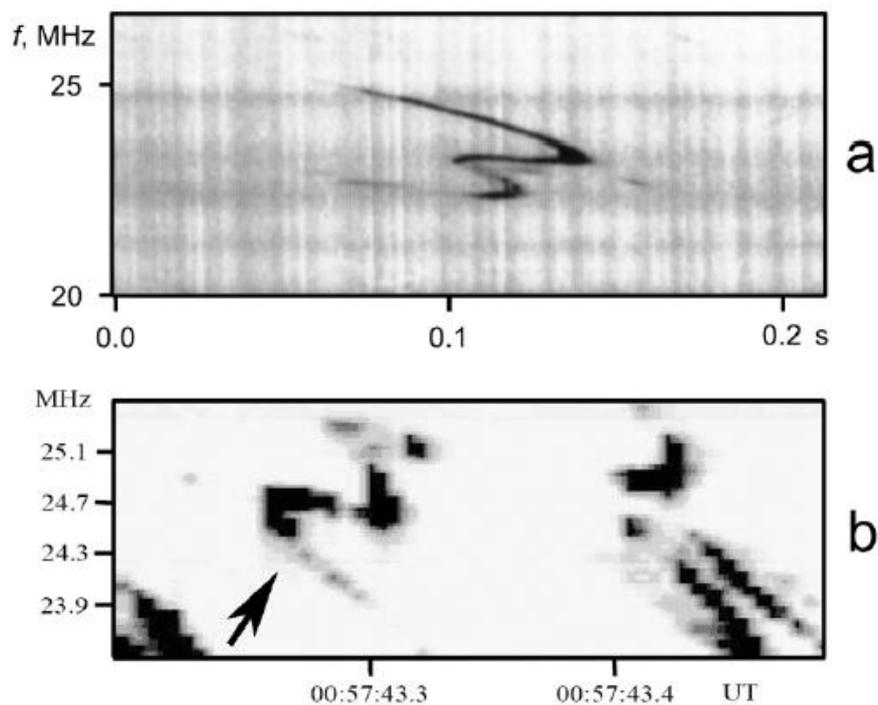

**Figure 3:** *Sometimes S-bursts show an ambiguity in frequency, when the emission from one radio source (i.e., from one emission band in dynamic spectrum) was observed at several frequencies simultaneously: **a)** 1988 November 19 (spectrum 106 in Riihimaa, 1992); **b)** 1995 May 16 (Ryabov et al., 1997).*

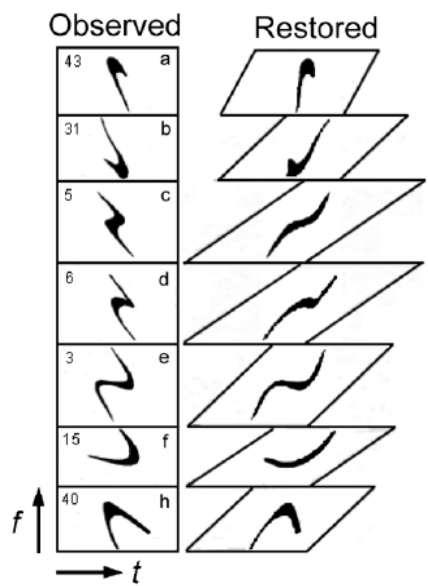

**Figure 4:** *The observed frequency-ambiguity in S-bursts (left column; plots and label numbers from Ryabov et al., 1997) could be transformed into non-ambiguity bursts (right column) using the correction for frequency-related delay. As a result of such idealized restoration, the rectangular frames in left plots became parallelograms in right.*

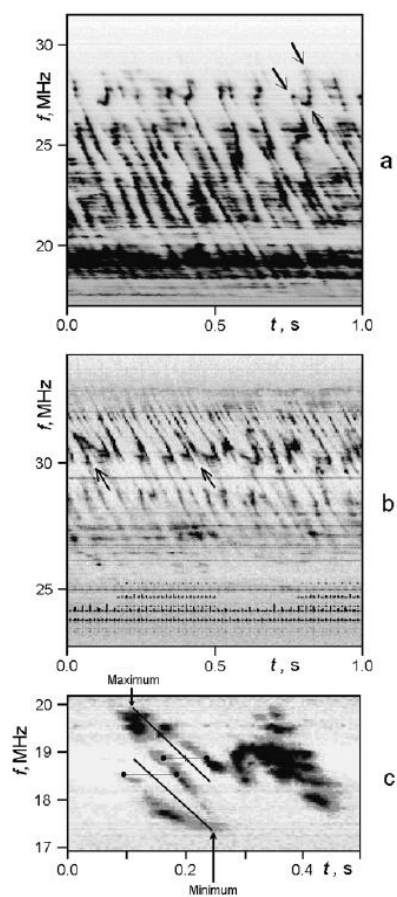

**Figure 5:** *The examples of frequency-ambiguity (arrowed) in S-burst trains: **a**) 1998 September 20 (UTR-2); **b**) 1995 May 16 (NDA); **c**) the use of middle lines (solid straight lines) to measure the distortion drift rate of such events (1995 May 16; UTR-2).*

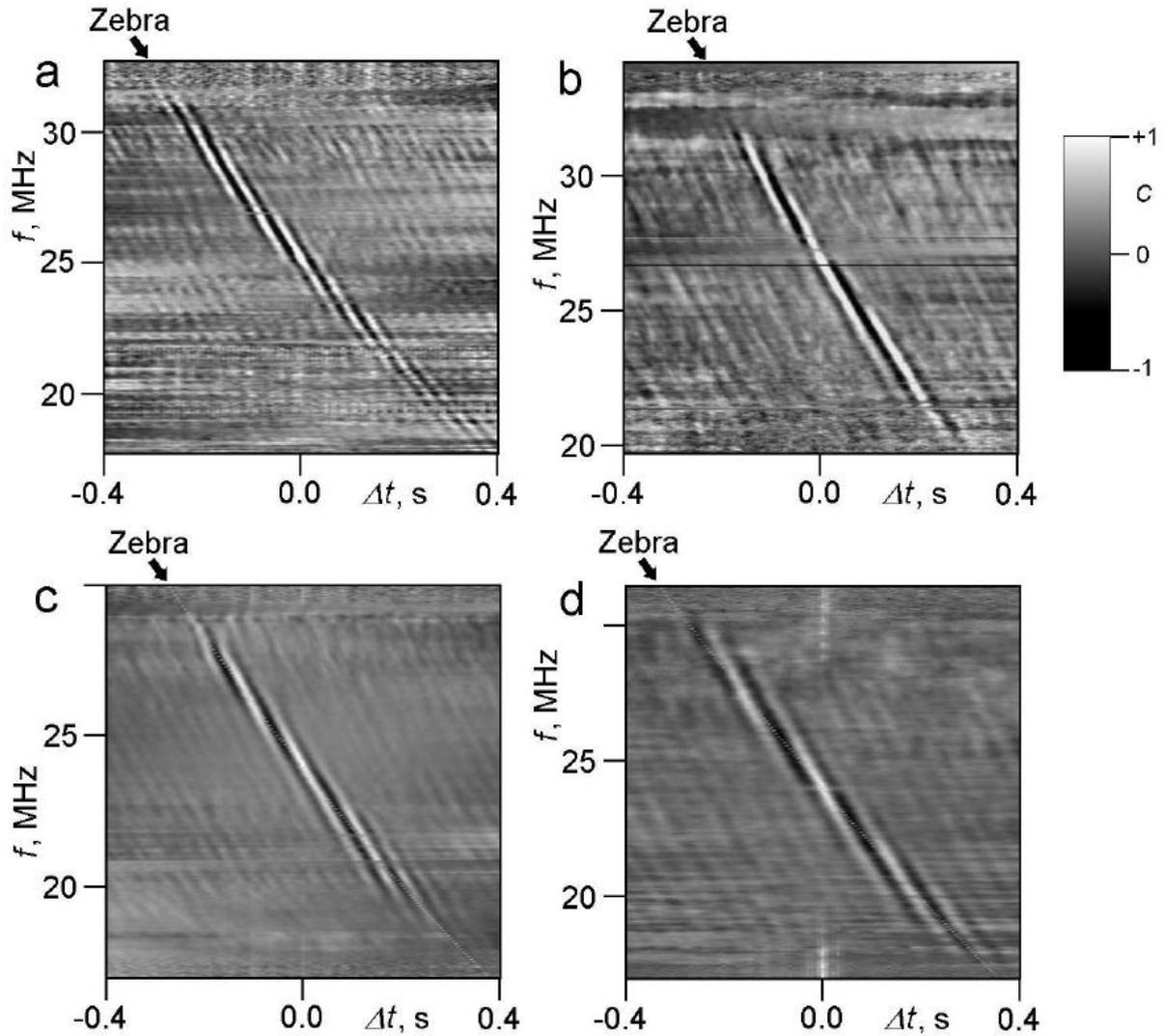

**Figure 6:** *The examples of correlation patterns of S-burst trains which were observed using different radio telescopes and equipment sets: (a) 1995 April 14 (NDA); (b) 1996 June 11 (NDA); (c) 1997 June 08 (UTR-2); (d) 1998 September 20 (UTR-2). The center lines of zebras are shown as white dotted lines in (c) and (d). The correlation is defined with the gray scale.*

$D_d = (f_e - f_m)/(t_e - t_m)$, where $f_e$ is the minimal or maximal frequency in the spectral feature; $f_m$ is the selected frequency for median searching; $t_e$ is the time of $f_e$ recording; $tm$ is the central time of the event at the selected frequency $f_m$. Of course, this manual method can lead to a large scatter of individual estimates of the distortion drift rate. However, the histogram can help reveal the most probable and correct value of $D_d$.

How does this spectral distortion manifest in the dynamics of regular S-bursts? It is reasonable to compare the measured value of spectral distortion $D_d$ and the common S-burst drift. To extract the average drift from a mix of many S-bursts, we apply the correlation method (Appendix). The essence of the method is the calculation of the linear correlation coefficient between the DAM spectral intensity in the current ($f$) and the reference ($f_o$) frequency channels for the used set of $\Delta t$ time shifts. Then details of the correlogram were visualized using a logarithmic scale and the normalization in each frequency channel of the spectrum analyzer (see Appendix). The typical visualized correlograms are shown in Fig. 6.

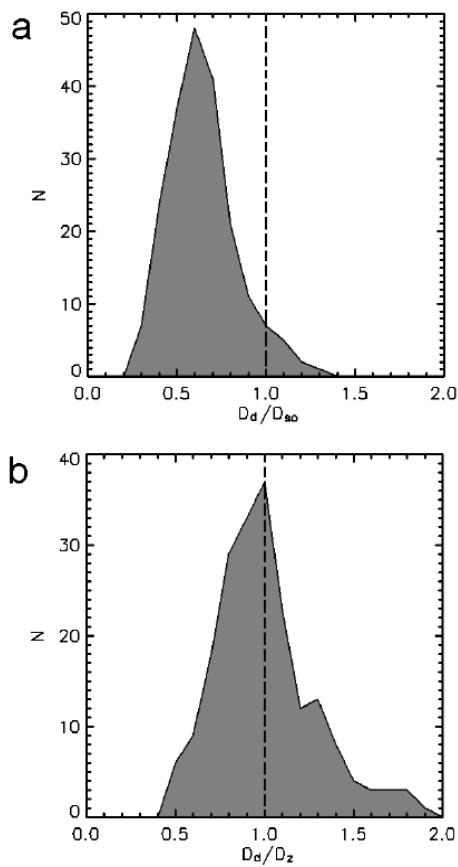

**Figure 7:** *The comparison between the distortion drift rate ($D_d$) of frequency-ambiguity events, like in Fig. 4, and the parameters of zebra correlation pattern like in Fig. 6: **a)** the histogram of $D_d/D_{so}$ estimates; **b)** the histogram of $D_d/D_z$ shows that $D_z$ can be used for $D_d$ measurement (i.e., for the distortion drift rate).*

One can see that the correlation patterns clearly show the system of drifting bright and dark bands, i.e., regions of correlation or anti-correlation respectively, which is named "zebra". Such zebras were described and modeled in our previous paper (Arkhypov and Rucker, 2012).

It is important to emphasize that the zebra patterns were detected using different equipment sets. Our processed data were recorded in 1994-1999 using different acousto-optical spectro-analysers with the largest decametric arrays in Nançay (NDA, Observatoire de Paris) and Kharkiv, Ukraine (UTR-2, IRA NASU). The UTR-2 data were recorded with the author's participation in 1995-1999 (Zarka et al., 1997). Moreover, the first zebras were revealed (Fig. 6 and 5 in Arkhypov and Rucker, 2009) using UTR-2 and the Digital Signal Processor (Konovalenko et al., 2001). Both antennae and 3 spectrum-analyzers confirm the reality of the zebra effect.

The frequency drift rates of individual bands in zebras were measured manually to estimate the average observed drift-rate of S-bursts ($D_{so}$) at different frequencies. As a result, we constructed the histogram *of $D_d/D_{so}$* ratios, selecting pairs of $D_{so}$ and $D_d$ estimates at approximately the same frequencies (Fig. 7a). One can see that the most probable ratio is about $D_d/D_{so} = 0.6$. The average ratio is $<D_d/D_{so}> = 0.64\pm0.02$ for 102 estimates in 6 spectra.

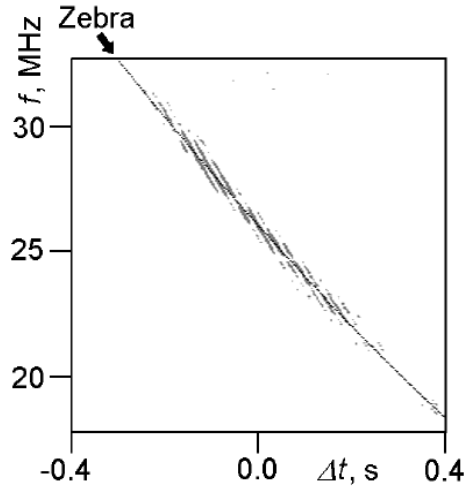

***Figure 8:*** *The example constructing of the central line for the zebra from Fig. 6a. The gray color marks regions in the correlogram with |C| > 0.99 (Appendix). The central line (solid curve) is the quadratic approximation or regression for the grey pixels.*

This means that the spectral distortion is not a sole factor controlling the S-burst drift rate. However, its contribution is significant because $D_d/D_{so} \sim 1$.

Another parameter of the zebra in correlogram is the frequency drift rate $D_z$ of its central line (Fig. 8). In fact, this line tracks the regions of maximal correlation or anti-correlation. In the case of broadband S-bursts (as in Fig. 2), the zebra is a result of deviations from the ideal quasi-linear burst. Hence, its drift rate $D_z$ displays the dynamics of perturbations in S-bursts. The central line of a zebra was found automatically using the regions with |C| > 0.99 (Appendix) in correlogram to find their approximation

$$f = f_o + a\,\Delta t + b\,\Delta t^2, \tag{1}$$

where $a$ and $b$ are constants found with the least-squares method. We have processed 32 dynamic spectra of broadband S-burst trains of 1995-1999 with clear zebra-patterns in their 2D-correlation plots. All spectra were recorded during Io-B storms (Fig. 9). S-bursts appeared or were recorded in different frequency bands in the presence of interference. To obtain the clear zebra-pattern, the reference frequency ($f_o$) was tuned for individual spectra. For a comparison and averaging of homogeneous data, the individual estimates of the coefficient a were reduced to the standard reference frequency $c_r = 25$ MHz as

$$a_r = -[a^2 - 4b(f_o - c_r)]^{1/2}, \tag{2}$$

The second coefficient is the same for all frequencies, so $b_r = b$. The reduced values $a_r$ and $b_r$ are plotted in Fig. 10. No significant correlation is revealed between these parameters. Formally, the correlation coefficient is $-0.26 \pm 0.16$. One can see that these individual estimates are clustered around the average values: $<a_r> = -20.0 \pm 0.4$ MHz/s and $<b_r>=+4.8\pm0.6$ MHz/s$^2$. The standard deviations of ar and br are 2.27 MHz/s and 3.13 MHz/s2 respectively. The scattering of the $a_r$-estimates rather exceeds the standard error ($\approx 1$ MHz/s). However, the variability of the parameter $b_r$ is not clear because its standard error is about its standard deviation. We use Eq. (1) and the obtained individual estimates $a$, $b$, and $f_o$ for calculation of frequency drift rate of the zebra central line

$$D_z = -[a^2 - 4b(f_o - f)]^{1/2}, \qquad (3)$$

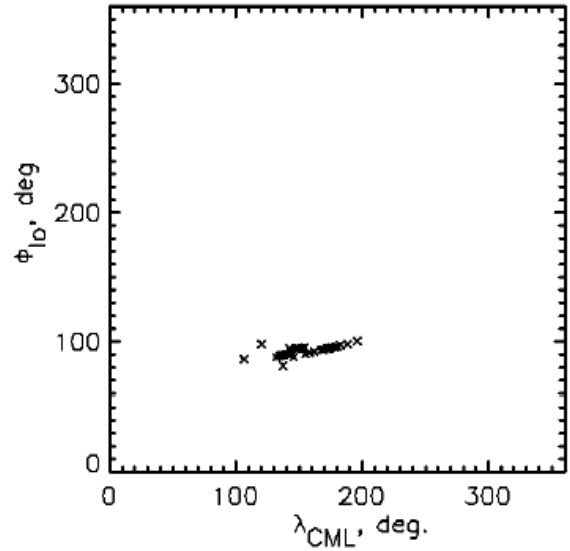

***Figure 9:*** *The selected records with broadband S-burst trains in the $\Phi_{Io} - \lambda_{CML}$ plot (crosses). Here $\Phi_{Io}$ is the orbital longitude of the jovian satellite Io. Another coordinate $\lambda_{CML}$ is the planetographic longitude of the central meridian (CML) on the Jupiter's disk, i.e., this is a longitude of an observer.*

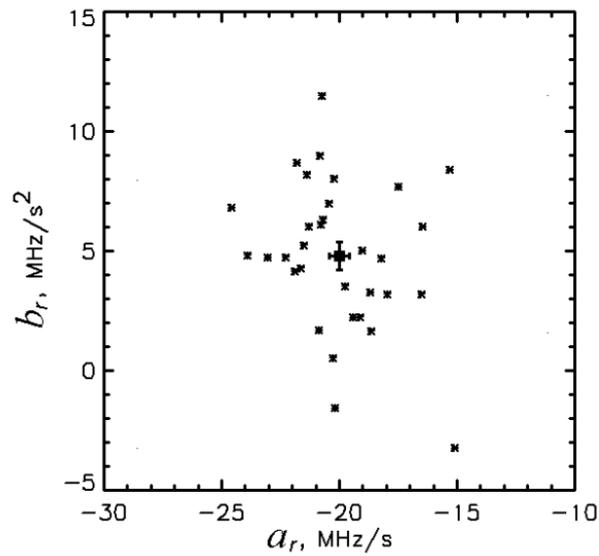

**Figure 10:** The estimates of coefficients in approximation of center lines of a zebra ($f = f_o + a\Delta t + b\ \Delta t^2$). The square with standard error bars is the cluster center.

Substituting into Eq. (3) the frequency $f$ of $D_d$ estimate, we can find the $D_d/D_z$ ratio. The histogram of such estimates is shown in Fig. 7b. One can see that the most probable value of $D_d/D_z$ is 1. The average value is $<D_d/D_z> = 1.01 \pm 0.03$ for 102 estimates in 6 spectra.

The obtained result $D_d = D_z$ means that the distortion drift rate ($D_d$) practically coincides with the drift rate of perturbations in usual S-bursts ($D_z$). This correspondence allows us to consider the center line of a zebra as the dispersion-like phenomenon also, which indicates that even ordinary S-bursts are distorted. Secondly, the zebra parameter $D_z$ can be used for $D_d$ measurement (i.e., the drift rate of spectral distortion). This is important, because the frequency-ambiguity is rare (< 1%) or absent in S-trains, nevertheless $D_z$ could be measured.

### 3. Analysis of corrected spectra

If the dynamic spectrum of S-bursts is distorted, as it is shown in the previous section, it makes sense to analyze the possible consequences of corrected dynamics of radio sources. The above estimates of the dispersion drift rate can be used for the correction of the observational time t at arbitrary frequency f to the receiving time scale to at $f_o$ reference frequency

$$t_o = t + \Delta t, \qquad (4)$$

where the correction $\Delta t = -a - [a^2 - 4b(f_o - f)]^{1/2}/(2b)$ is the solution of Eq. (1). The plot $F(t_o, f)$ is the S-burst storm as it could be seen without the dispersive distortion (Fig. 11).

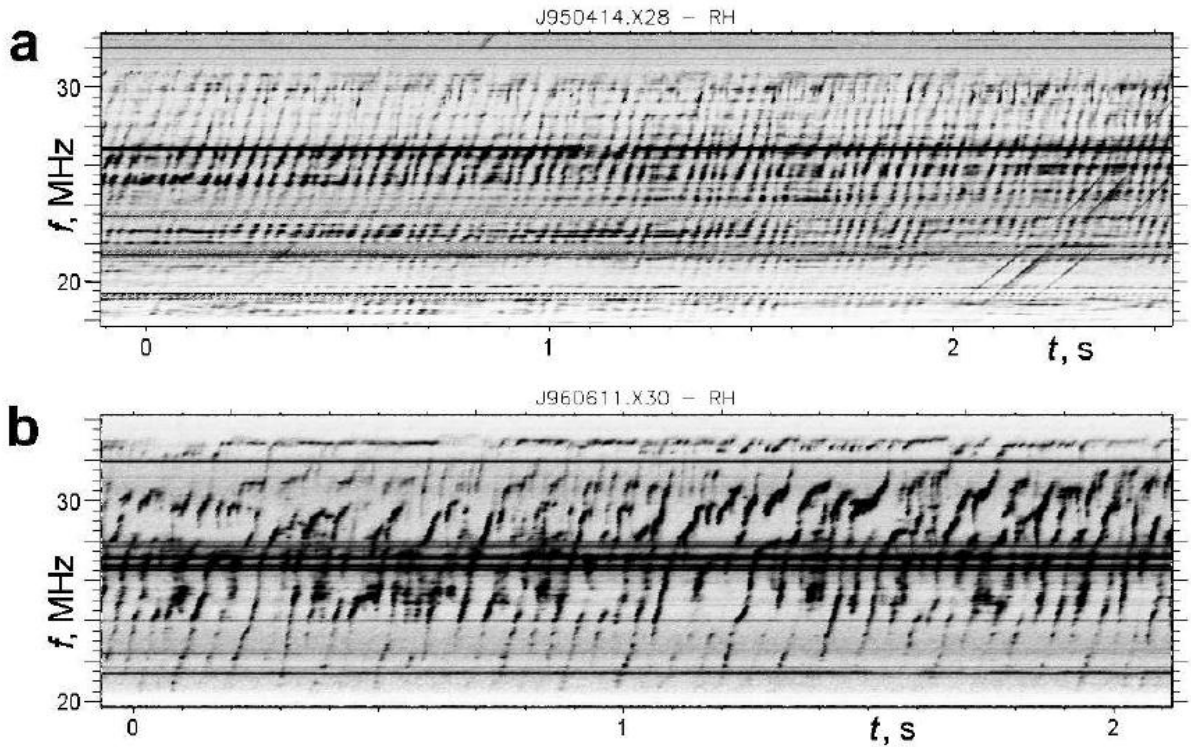

**Figure 11:** *The same spectra from Fig. 2, but after our time-delay correction.*

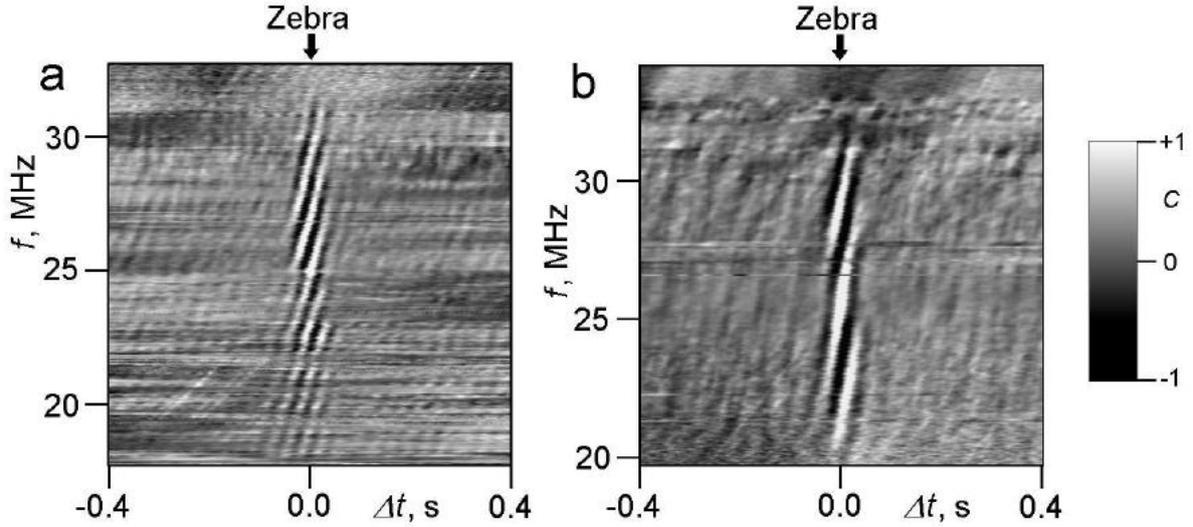

**Figure 12:** *The correlation patterns of corrected spectra in Fig. 11: (a) the quasi-linear S-bursts (NDA; 1995 April 14, 05:43 UT); (b) the perturbed S-bursts (NDA; 1996 June 11, 23:40 UT).*

The important feature of corrected spectra is the transformation of negative frequency drift of S-bursts (Fig. 2) into a mainly positive drift (Fig. 11). The averaged reverse drift of S-bursts is shown as the positive frequency drift of the pattern in the zebra-band of correlogram of the corrected spectrum (Fig. 12). This formally means that the radio sources in an S-burst train do not move away from Jupiter, as traditionally thought, but they fly in opposite direction, toward the planet.

There is a simple interpretation of this correlogram. As shown earlier (Arkhypov and Rucker, 2012), a zebra-band is reproduced as a result of perturbed motion of several radio sources in the parallel electric field of a standing Alfvén wave. It is known that any standing wave is described as the product of temporal and spatial terms with independent phases. The temporal phase of the standing wave changes synchronously at all altitudes. This synchronism is manifested as a vertical zebra-region in Fig. 12 after transformation of the altitude scale into the corresponding gyrofrequencies of electrons.

The manual measurements of frequency drift rate of bright or dark bands in 32 zebras of 1995-1999, like in Fig. 12, are summarized in Fig. 13. Averaging over the groups of four individual cross-estimates (squares with rms error bars) reveals an increasing in the corrected drift rate of S-bursts ($D_{sc}$) with frequency.

For comparison, we calculated the theoretical curves based on the opinion that electrons moves mainly adiabatically and generate S-bursts through the cyclotron maser instability at a loss cone boundary (Zarka et al., 1996). The frequency drift rate of an S-burst ($D_{sc}$) is expressed by the velocity ($V_∥$) of the radio source motion along the magnetic field,

$$D_{sc} = \kappa V_∥,\qquad\qquad(5)$$

where $V_∥ = [V^2 - V_⊥^2]^{1/2}$; $V$ is the total electron velocity (labeled in Fig. 12); $V_⊥ = V \sin \alpha$; $\alpha$ is the pitch angle between the vector of electron velocity and the magnetic field. In the case of adiabatic motion in an inhomogeneous magnetic field, an electron preserves the invariant $\sin^2\alpha/f_{ce}$, where $f_{ce}$ is the local gyrofrequency of electrons (Hess et al., 2007a). The parameter $\kappa \approx 10^{-6}$ MHz/m is the typical gradient of the electron gyrofrequency at DAM sources. Since the maser works at the frequency $f \approx f_{ce}$ in the low-density plasma near Jupiter, the pitch angle can be found as

$$\sin \alpha = [f/f_m]^{1/2}, \qquad (6)$$

where $f_m$ is the maximal gyrofrequency of electrons in the jovian ionosphere, where some of the electrons precipitate into the atmosphere forming the loss cone.

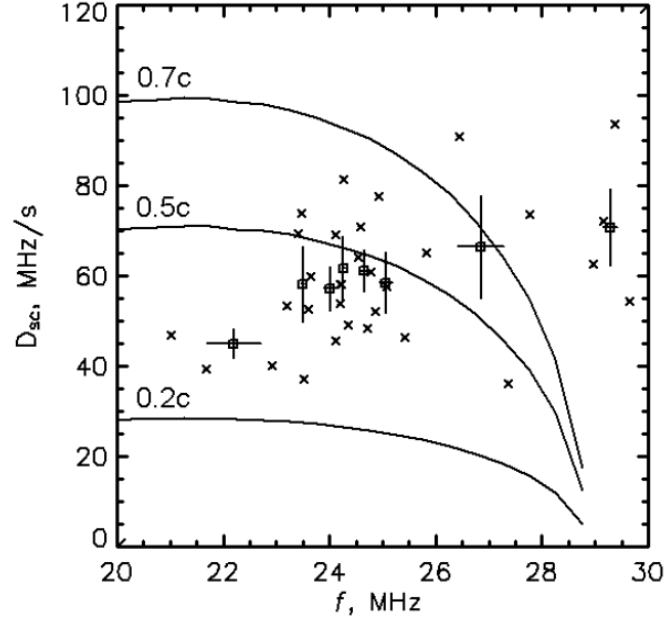

***Figure 13:*** *The estimates (crosses) of the corrected drift rate of S-bursts ($D_{sc}$) are found as the frequency drift rate of white/black strips in correlation zebras (like in Fig. 12) which are constructed using corrected spectra (like in Fig. 11). The open squares with error bars are the averages for 4 crosses. For comparison, the theoretical curves are calculated for electrons traveling adiabatically toward Jupiter at the loss cone boundary (where S-burst generation is believed) with the labeled total electron velocity. All curves are based on VIP-4 magnetic field model (Connerney et al., 1998) and the leading angle of $20^o$ for S-bursts in Io-B storms (Ryabov, 1994).*

Using VIP-4 model of jovian magnetic field (Connerney et al., 1998), we have tracked the magnetic field line, which intersects the average effective position of Io satellite (longitude $222^o$) during the generation of the observed S-storms. At that the true mean Io longitude has been reduced by $20^o$ to take into account the known effect of the lead angle for S-bursts (Ryabov, 1994). The minimal value of $f_m=28.9$ MHz at the southern footprint of this magnetic line is the cutoff frequency of the theoretical curves. As a result, one can see the disagreement between adiabatic approximation and experimental $D_{sc}$ in Fig. 13. Hence, the field aligned potentials or dispersive Alfvén waves could significantly disturb the dynamic of S-burst sources (Hess et. al, 2007a; Ergun et al, 2006).

The average drift rates after correction is $<D_{sc}> = 59.8 \pm 2.6$ MHz/s. Using the cyclotron frequency gradient $\kappa= 1.14$ kHz/km (VIP4 magnetic model), this value corresponds to the velocity $<D_{sc}>/\kappa = (0.17 \pm 0.01)c$.

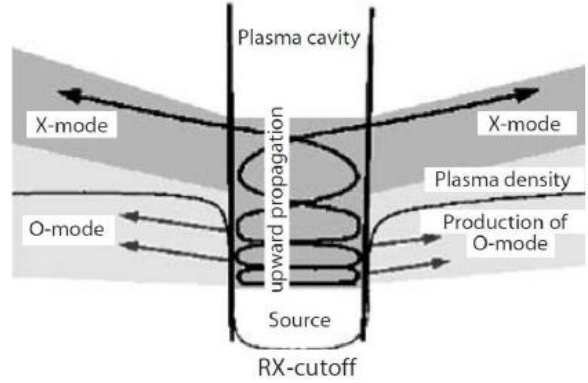

**Figure 14:** *Schematics of a cavity which traps the electron-cyclotron maser radiation of AKR (Treumann, 2006). The radio emission escapes the cavity mainly as X-mode at the level where its frequency f is just above $f_x$ of the surrounding dense plasma (solid curve labeled "RX-cutoff").*

## 4. Discussion

Our phenomenological analysis reveals that S-bursts in DAM dynamic spectrum are distorted (Sect. 2). The mentioned distortion manifests as a regular decrease in the receiving time of radiation with increasing frequency. Formally, this is a dispersion-like effect. The alternative interpretation in terms of some modulator of S-burst generation (Arkhypov and Rucker, 2011a), which drifts with velocity $D_d$, principally cannot reproduce the spectral features with ambiguity in frequency (Figures 2-4).

However, the value of found distortion far exceeds the expectations of usual dispersion. In the high-frequency approximation of emission delay in the interplanetary medium is $(L/c)(1-f_{pe}^2/f^2)^{-1/2}$, where $L \sim 5$ AU $= 7.5 \times 10^8$ km is the typical Jupiter-Earth distance; $f_{pe} \sim 20$ kHz is the typical plasma frequency. Hence, the usual difference of radiation delays between frequencies of 20 MHz and 30 MHz is only $\sim 7 \times 10^{-4}$ s. However, the zebra in Fig. 6 can be tracked up to 0.7 s in the time lag axis. In principle, such large delay difference is possible near the X-mode cutoff frequency $f_x$, where the DAM generation is believed (Zarka, 1998). The X-mode group velocity near the cutoff is given in Hewitt and Melrose (1983)

$$V_g/c = n_x[(1 - Y) / (2 - Y)](1 + 3 \cos^2\theta)^{1/2}, \tag{7}$$

where $n_x$ is the refractive index; $Y = f_{ce}/f$; $f_{ce}$ is the electron gyrofrequency; $\theta$ is the angle between the magnetic and wave vectors. As $n_x \to 0$ at $f \to f_x$ (Fig. 2 in Hewitt and Melrose, 1983), the group velocity formally can be very small. Using Eq. (7) in our simulation, we have obtained the formally realistic frequency dependence of $D_d$ in much of the frequency range of S-bursts from 6 MHz to 25 MHz (Arkhypov and Rucker, 2012). The deviation of the calculated curves from the experimental points at $f > 25$ MHz means that our modeling needs to be improved.

Another problem is that the high stability of the ratio $f/f_{ce}$ is needed for the relative stability of the distortion coefficients in Fig. 10. Moreover, in general the cyclotron maser generation is possible in two frequency bands with different $f/f_{ce}$ and group velocities (Hewitt and Melrose, 1983; Willes, 2002). This problem could be resolved in the cavern model of aurora kilometric radiation of the Earth (AKR; Fig. 14). It is found on a base of measurements in situ that AKR is generated inside depleted cavities with tenuous and hot plasma at the frequencies below $f_{ce}$ and $f_x$ of the surrounding cold plasma (Treumann, 2006). That is why the radiation can leave the narrow cavity only at certain distance above the radio source, where the cutoff frequency outside the cavity decreases just below the radiation frequency $f$. Hence, this escape condition fine-tunes the $f_{ce}$ and stabilizes the ratio $f/f_{ce}$. Note that Jovian S-bursts also indicate the signs of intra-cavern origin (Arkhypov and Rucker, 2012).

To show that this approach is promising, we numerically calculate the dispersion drift rate of DAM, using the realistic model. We suppose that the S-burst radiation escapes the cavern where the following condition is realized

$$f = f_x + A \, \varepsilon \, f_{ce}, \tag{8}$$

where $f_x = f_{ce}(1+\varepsilon)$ is the cutoff; $\varepsilon = (f_{pe}/f_{ce})^2$; $f_{pe}$ and $f_{ce}$ are the plasma and cyclotron frequencies of electrons; $A$ is the constant scale factor. We accept $A$=0.5 as a formal compromise between a ray deflection from the Earth, because the great refraction ($n_x << 1$) at $f \to f_x$, and the vanishing of the dispersion effect at $f - f_x >> \varepsilon f_{ce}$.

Using Eq. (8) we calculated the dispersion drift rate for the X-mode propagating from the radio source along the straight line to the Earth. Since S-bursts are observed at angles up to $90^o$ to the magnetic field in the radio source (Ryabov, 1994), we consider the ray which propagates toward the Earth, perpendicularly to the magnetic field in the source, but in parallel with the jovian equator plane (the Earth declines from this plane less than $3.3^o$). Then the dispersion drift rate in frequency is

$$D_d = (f_2 - f_1) \, / \, [\delta t(f_2) - \delta t(f_1)], \tag{9}$$

where $f_2 - f_1 = 2$ MHz; and the emission delay is

$$\delta t(f) = \int_{\Lambda}^{0.5 R_J} ds / V_g, \tag{10}$$

where $ds$ is the path element along the ray; $\Lambda$ is the ray path length in the cavity. We calculate the group velocity with the general equation for the wave propagation in cold plasma (Hewitt and Melrose, 1983):

$$V_g = \frac{c n_x}{n_x \partial (f n_x)/\partial f} \sec(\theta - \theta_x), \tag{11}$$

where $c$ is the light velocity; $\theta$ is the angle between the magnetic and wave vectors; $\theta_x$ is the angle between the magnetic vector and the group velocity direction of the wave. It follows from Eqs. A4, A9 and A10 in Hewitt and Melrose (1983) that $\theta - \theta_x \to 0$ when $\theta \to \pi/2$. The derivative in Eq. (11) was calculated numerically using the small variation in frequency with the step of $0.01(f - f_x)$.

The refractive index $n_x$ of X-mode was calculated with the Appleton-Hartree equation:

$$n_x^2 = 1 - \frac{X(1-X)}{1 - X - 0.5Y^2 \sin^2\theta - \sqrt{0.25Y^4 \sin^4\theta + (1-X)^2 Y^2 \cos^2\theta}}, \tag{12}$$

where $X = (f_{pe}/f)^2$. To calculate the wave delay in Eq. (10), we used numerical integration in steps of 0.001 $R_J$, where $R_J$ is the radius of Jupiter. Since the main contribution to the delay of the emission is formed at a distance of $< 0.1 R_J$ from the radio source, the upper limit of the integration is chosen at a sufficient formal distance of $0.5 R_J$. We used VIP-4 model of jovian magnetic field (Connerney et al., 1998) and the average longitude of radio sources of $179^o$ due to underlying data.

The electron number density in the source region for fpe calculation is selected as a combination of known components

$$N_e = N_o[K_1 \exp(-\frac{s}{h_1}) + K_2 \exp(-\frac{s}{h_2}) + K_3 \exp(-\frac{s}{h_3}) + K_4],$$

(13)

where $s$ is the length of path along the magnetic field from the ionosphere peak up to the northern radio source; $K_i$ are the contribution coefficients of the components number $i$ in the total electron number density $N_o$ at the ionosphere peak; hi is the scale height of the component. Using the radio occultation data (Strobel and Atreya, 1983; Yelle and Miller, 2004), we assumed the quite realistic number density at the morning ionosphere peak $N_o = 3 \times 10^4$ cm$^{-3}$ and the typical scale height of 800 km. Unfortunately, the electron number density was not measured in the regions of DAM sources. That is why we have chosen plausible values of the model parameters (Table 1) so as to demonstrate the dispersion possibility.

**Table 1:** Components of the electron profile in the dispersion model

| Component (i) | Portion ($K_i$) | Scale height ($h_i$), $R_J$ | Interpretation |
|---|---|---|---|
| 1 | 0.621 | 0.011 | Cold ionosphere ($kT = 0.1$ eV)* |
| 2 | 0.31 | 0.05 | Warm ionosphere at aurora ($kT = 0.5$ eV) |
| 3 | 0.068 | 0.154 | Polar wind ($kT = 1.5$ eV) |
| 4 | 0.00017 | – | Hot electrons ($kT \gtrsim 50$ eV) |

(*) $kT = mgH$, where m is the half of the proton mass in a hydrogen plasma; $g$ is the gravity acceleration; $k$ is the Boltzmann constant.

Figure 15 shows the results of our calculations. The accordance of the calculated curve with experimental points shows that the dispersion can create the observed distortion of the DAM spectra. The reproduction of decreasing of the dispersion curve at high frequencies is especially interesting. This effect of the cold ionosphere is an unknown alternative for the adiabatic interpretation.

The association of AKR and DAM radiations with magnetospheric cavities does not exhaust the list of analogies. For example, the negative frequency drift of narrow-band AKR-bursts is traditionally interpreted assuming that the radio sources drifted upward (Treumann, 2006). Hence, the dispersion is believed to be insignificant. However, Ergun et al. (2006) note: "The AKR spectra observed inside of the auroral cavity do have fine structure, but the frequency drifts are both positive and negative and indicate slower source speeds... The spectra in Figure 4d is qualitatively different from those observed inside of auroral cavities." This difference between inside and outside dynamics could be interpreted in favor of a strong dispersion effect.

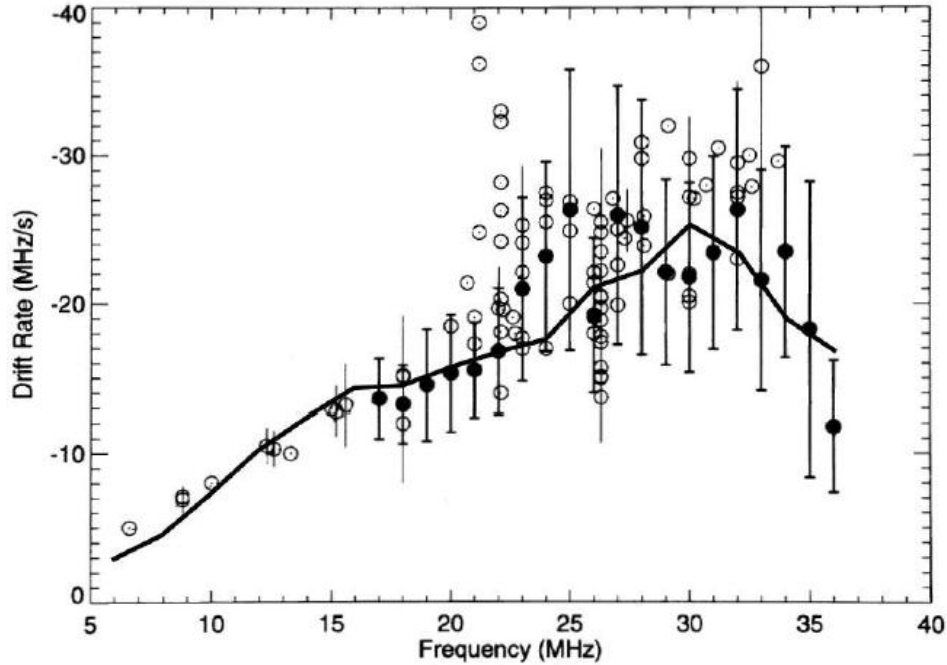

**Figure 15:** *The observed drift rates of S-bursts (Zarka et al., 1996) are in accordance with our model curve for the dispersion effect, which is calculated using the realistic electron profile in Eq. (13) and the cavern model (Fig. 14) of the radio source.*

Satellite measurements indicate in-situ a relationship between AKR and the downward electron flows in the cavity (e.g., Zarka 1998; Treumann, 2006). The positive frequency drift of Jovian S-bursts, found after our correction of spectral distortion, shows the downward source motion, too. These intriguing Earth-Jupiter parallels show that the dispersive distortion in S-burst dynamics is worth of further research.

## 5. Conclusions

There are experimental facts which do not support the traditional believe in negligible dispersion effect in the dynamic spectra of Jovian S-burst emission.

1. It was found that up to 1 % of bursts in an S-burst train are of distorted meandering shape in dynamic spectrum, where the emission from one radio source was observed at several frequencies simultaneously (Figs. 3-5). It is not easy to explain such spectra in terms of causality. However, this problem vanishes if the frequency-depended distortion of the spectrum is supposed.

2. The frequency drift rate of middle lines of such frequency-ambiguity events statistically coincides with the drift rate of perturbations in common S-bursts (Fig. 7b). This indicates a general distortion of the dynamic spectrum of common S-bursts.

3. We did not find clear peculiarities in the frequency dependence of the spectral distortion. This distortion is the broadband effect with the regular dependence of frequency. Hence, we cannot confirm the significant role of local electric potential structures in the forming of S-bursts with ambiguity in frequency.

4. The simulation shows that the dispersion is capable in principle to reproduce the found spectral distortion.

5. The correction of the spectral distortion reverses the drift of S-burst sources from negative to positive. The corrected dynamics of S-bursts does not support the adiabatic approximation of the motion of emitting electrons. Hence, the recognition of significant dispersion effect in S-burst emission could lead to a revision of the traditional opinions on S-burst dynamics.

Anyway, the dispersion-like phenomena in Jovian S-bursts deserve discussion because they have no satisfactory explanations in terms of traditional approach.

**Acknowledgements.** The authors thank P. Zarka (LESIA, Observatoire de Paris) and V.B. Ryabov (IRA NASU) for their donation of some Nançay and UTR-2 DAM records for these studies.

**Appendix A.** The correlation method

We construct a matrix of estimates of the linear correlation coefficient r between the spectral intensity F in the current (**f**) and the reference ($f_o$) frequency channels for used set of $\Delta t$ time shifts. These correlation estimates are calculated for a whole DAM dynamic spectrum, i.e, for every available frequency channel taking into account the whole set of $F$ readings. The following formula was used for the processing of a pixel image of spectrum

$$r(i, \Delta k) = \sum_{k=\{^{1, \Delta t \geq 0}_{1-\Delta k, \Delta t < 0}}^{\{^{m-\Delta k, \ \Delta k > 0}_{m, \ \Delta k < 0}}} \frac{[F(i_o, k) - \langle F \rangle_{i_o}][F(i, k + \Delta k) - \langle F \rangle_i]}{M \sigma_{i_o} \sigma_i},$$

(A.1)

where: $i$ and $k$ are the pixel numbers in the frequency and time scales; $i_o$ is the line number corresponding to the reference frequency $f_o$; m is the width in pixels of the analyzed dynamic spectrum; $\Delta k$ is a time shift in pixels; $<F>_i$ is the average spectral intensity along the spectral line at constant i or radio frequency; $M = m - |\Delta k|$ is the number of terms in the sum; $\sigma_{i_o}$ and $\sigma_i$ are the standard deviations of the spectral intensity at the corresponding lines. For example

$$\sigma_i = \sqrt{\sum_{k=\{^{1, \Delta t \geq 0}_{1-\Delta k, \Delta t < 0}}^{\{^{m-\Delta k, \ \Delta k > 0}_{m, \ \Delta k < 0}} \frac{[F(i, k + \Delta k) - \langle F \rangle_i]^2}{M - 1}},$$

(A.2)

Such calculations were performed for each pixel on the plot $i-\Delta k$. The corresponding steps in frequency and time are 15 kHz to 50 kHz and 4 ms to 5 ms. We used the negligible shifts ($|\Delta t| \leq 0.4s$) compared to the duration of the spectrum (from 22 s to 89 s or 6,000 $\leq m \leq$ 18,000). The estimates $r(i, \Delta k)$ or $r(f, \Delta t)$ were coded in the colors with a gray linear scale to be analyzed as an image (e.g., Fig. 6).

The correlation coefficient between the spectral intensities in f and $f_o$ frequency channels with the _t time shift is usually $r(f, \Delta t) < 0.1$ with the exception of the sharp peak $r \approx 1$ at $\Delta t \approx 0$ and $f \approx f_o$. To show the whole correlation pattern in details, some logarithmic parameter is more appropriate than linear $r(f, \Delta t)$. As $-1 \leq r(f, \Delta t) \leq +1$, we use the logarithmic parameter in the form

$$Y(f, \Delta t) = \lg[r(f, \Delta t) + 1], \qquad (A.3)$$

Moreover, interferences appear as numerous artifacts in the form of horizontal stripes in $r(f, \Delta t)$-plots. To standardize the average brightness of a line in the plot, we use the difference between $Y(f, \Delta t)$ and its average value in this frequency channel $<Y>_f$. To standardize the amplitude of variations in the difference, we normalize the difference with its standard deviation calculated in the same frequency channel.

Following these considerations, we visualize the whole correlation pattern using the parameter $C(f, \Delta t)$

$$C(f, \Delta t) = \frac{Y(f, \Delta t) - \langle Y \rangle_f}{3\sigma_f}, \qquad (A.4)$$

where $Y(f, \Delta t)$ is the logarithmic parameter according to Equation A.3; $<Y>_f$ and $\sigma_f$ are the average $Y(f, \Delta t)$ value and its standard deviation for each frequency channel.